\documentclass[manuscript,screen]{acmart}

\AtBeginDocument{%
  \providecommand\BibTeX{{%
    \normalfont B\kern-0.5em{\scshape i\kern-0.25em b}\kern-0.8em\TeX}}}

\setcopyright{acmcopyright}
\copyrightyear{2022}
\acmYear{2022}
\acmDOI{XXXXXXX.XXXXXXX}

\acmConference[FAccTRec2022]{5th FAccTRec Workshop on Responsible Recommendation at RecSys 2022}{September, 23,
  2022}{Seattle, WA, USA}
\begin{document}

\title{The Role of Bias in News Recommendation in the Perception of the Covid-19 Pandemic.}

\author{Thomas Elmar Kolb}
\email{thomas.kolb@tuwien.ac.at}
\orcid{0000-0002-2340-0854}
\author{Irina Nalis}
\email{irina.nalis-neuner@tuwien.ac.at}
\orcid{0000-0001-7101-3229}
\author{Mete Sertkan}
\email{mete.sertkan@tuwien.ac.at}
\orcid{0000-0003-0984-5221}
\author{Julia Neidhardt}
\email{julia.neidhardt@tuwien.ac.at}
\orcid{0000-0001-7184-1841}
\affiliation{%
  \institution{Christian Doppler Laboratory for Recommender Systems, TU Wien}
  \streetaddress{Favoritenstraße 9-11/194-04}
  \city{Vienna}
  \state{Vienna}
  \country{Austria}
  \postcode{43017-6221}}

\renewcommand{\shortauthors}{Kolb, et al.}

\begin{abstract}
  News recommender systems (NRs) have been shown to shape public discourse and to enforce behaviors that have a critical, oftentimes detrimental effect on democracies. Earlier research on the impact of media bias has revealed their strong impact on opinions and preferences. Responsible NRs are supposed to have depolarizing capacities, once they go beyond accuracy measures. We performed sequence prediction by using the BERT4Rec algorithm to investigate the interplay of news of coverage and user behavior. Based on live data and training of a large data set from one news outlet ``event bursts'', ``rally around the flag" effect and ``filter bubbles'' were investigated in our interdisciplinary approach between data science and psychology. Potentials for fair NRs that go beyond accuracy measures are outlined via training of the models with a large data set of articles, keywords, and user behavior. The development of the news coverage and user behavior of the COVID-19 pandemic from primarily medical to broader political content and debates was traced. Our study provides first insights for future development of responsible news recommendation that acknowledges user preferences while stimulating diversity and accountability instead of accuracy, only. 
\end{abstract}

\begin{CCSXML}
<ccs2012>
 <concept>
  <concept_id>10010520.10010553.10010562</concept_id>
  <concept_desc>Computer systems organization~Embedded systems</concept_desc>
  <concept_significance>500</concept_significance>
 </concept>
 <concept>
  <concept_id>10010520.10010575.10010755</concept_id>
  <concept_desc>Computer systems organization~Redundancy</concept_desc>
  <concept_significance>300</concept_significance>
 </concept>
 <concept>
  <concept_id>10010520.10010553.10010554</concept_id>
  <concept_desc>Computer systems organization~Robotics</concept_desc>
  <concept_significance>100</concept_significance>
 </concept>
 <concept>
  <concept_id>10003033.10003083.10003095</concept_id>
  <concept_desc>Networks~Network reliability</concept_desc>
  <concept_significance>100</concept_significance>
 </concept>
</ccs2012>
\end{CCSXML}

%\ccsdesc[500]{Computer systems organization~Embedded systems}

%\keywords{datasets, neural networks, gaze detection, text tagging}
\setcopyright{none}

\maketitle

\section{Introduction}

%% Was die arbeit selber macht, fehlt noch

``The virus is a democratic imposition'', was stated by the then German chancellor, Angela Merkel, in summer 2020. This statement resonated with many citizens and summarized the inseparable, multiple impacts of the Corona virus, with specific challenges for democratic societies concerning trust in institutions and political leadership \cite{schraff2021political}. Media outlets play a significant role in informing the public. Therefore, creation and use of news recommendations raise ethical, social, and legal difficulties and ask for a research agenda for socially responsible recommendation. User exposure to news articles on social media, news aggregation apps, and search engines is influenced by news search and recommendation and oftentimes builds on accurate predictions derived from former behavior. As summarized in Ekstrand et al. (2022, \cite{ekstrand2022fairness}) users' potential social and political decisions are influenced by recommendations. Furthermore, bias has a particular impact. It can result in incorrect or warped impressions of the world, which can have far-reaching effects. Therefore, bias mitigation techniques are essential. 

 Recommender systems are one of the most pervasive and most valuable machine learning applications in industry today \cite{meinl_recommender_2020}. The accuracy of these systems often is rather high, i.e., the recommendations made by the system are correct, but people are not really surprised by what is proposed to them \cite{jannach2016recommender}. It can be assumed that media biases \cite{eberl2017one} which lack in fairness and accountability might have played a role in changes of the perception of the pandemic. The exploration of beyond-accuracy measurements for fair recommender systems is necessary \cite{10.1145/3437963.3441665,abdollahpouri2021user} to create a responsible news recommender system that prevent the creation of "filter bubbles" and mitigates political polarization. Moreover, beyond accuracy measures could contribute to more diverse and fair coverage seem crucial to avoid the fostering of polarization through news recommender systems.
 
 However, until today much of the research on the role of media coverage and user behavior, only covers either very small data sets, short time spans and/or apply methods such as manual coding conducted without data scientists. What is especially missing is the analysis of temporal effects over time regarding the news coverage and user behavior around COVID-19. In this paper an exploration of changes in news coverage and user behavior is presented and the role of recommender systems is investigated. The aim is to unearth first insights on the role of bias in news recommendation in the perception of the COVID-19 pandemic.

To examine the temporal dynamics of the news coverage and user behavior on the global COVID-19 pandemic, three effects as described in the social sciences will serve as theoretical underpinning. First, event based "news bursts" and the relation to changes in topics will be explored. According to Bomberg et al. (2020)\cite{boberg2020pandemic} patterns of event based "news bursts" showed that reporting in the first weeks of the pandemic grew in tandem with the exponential spreading of the Corona virus. Secondly, the phenomenon of the ``rally around the flag" effect \cite{johansson2021rally} indicates that driven by acute fear individuals tend to flock around the fireplace of mainstream media outlets and which is supposed to show in changes of traffic. Thirdly, the "filter bubble" \cite{pariser2011filter} effect is known to potentially lead consumers to only be exposed to news stories that support their viewpoints, further polarizing society. 

Our contribution lies in the analysis of a large news data set which revealed the following outcomes.

\begin{itemize}
\item The Covid-19 pandemic affected a change in news coverage in line with the concept of "event bursts", hence a stark increase of coverage of the topic, an increase in the amount of coverage and a change in most popular featured topics are visible.
\item Online newsreaders' commenting behavior was affected by the pandemic in the direction of the "rally around the flag" effect, with a temporal dynamic that changed from a very strong increase in user activity which was followed by a steady decline.
\item The "filter bubble" effect became visible in disparities that were shwon through a model trained on sequence prediction. Thus, similar topics were recommended more frequently during the pandemic than before the pandemic, which shows the "filter bubble" effect through a decrease in diversity.
\end{itemize}

In the following sections, the theoretical and conceptual underpinnings of our research focus as well as our steps of data analysis and model training are outlined and the results are presented and discussed.

\section{Related Work}

 Since the outbreak of the global pandemic, several studies undertook investigations of the impact of the pandemic outside the medical realm and with a focus on the spheres of media. In the following section an overview of recent publications that partially cover phenomena in relation to the COVID-19 pandemic and showed aspects of event related "news bursts", "rally around the flag" effect or the development of "filter bubbles".

Regarding the phenomenon of even related "news bursts" \cite{boberg2020pandemic} and bias, earlier research on the impact of media bias has revealed the strong impact of media biases on opinions and preferences \cite{eberl2017one}.
%Moreover, it has been shown that trust in institutions and traditional media outlets are related to behavioral outcomes, such as the compliance with containment-measures was which were imposed by governments and required  unprecedented changes in the daily lives \cite{imhoff2020bioweapon}.
Moreover, it has been shown that trust in institutions and traditional media outlets are related to behavioral outcomes, such as compliance with containment measures imposed by governments and requiring unprecedented changes in daily life \cite{imhoff2020bioweapon}.
%The question on bias and agenda setting \cite{eberl2017one} was investigated for instance by Ng and Wen Tan (2021) \cite{ng2021diversity} who monitored the diversity of news media coverage in several countries in relation to cultural values, government stringency and pandemic severity.
For instance, Ng and Wen Tan (2021) \cite{ng2021diversity} addressed the question of bias and agenda setting \cite{eberl2017one} by monitoring of the diversity of news media coverage in several countries concerning cultural values, government stringency, and pandemic severity. They compared media diversity with cultural values and showed an increase in media diversity in higher collectivist cultures.

In addition, research on the social psychological effect of ``rally around the flag" effect demonstrates the need for understanding the temporal effects of media coverage on behavior. Earlier research on the impact of media bias has revealed the strong impact of media biases on opinions and preferences \cite{eberl2017one}. Moreover, it has been shown that trust in institutions and traditional media outlets are related to behavioral outcomes, such as the compliance with containment-measures which were imposed by governments and required  unprecedented changes in the daily lives \cite{imhoff2020bioweapon}. This showed in compliance which already in earlier research has been shown to follow systematic patterns of steep rises to rapid declines over time \cite{schraff2021political,johansson2021rally}. Thus, it can be assumed that in parallel to changes in user behavior regarding media consumption as described in the "rally around the flag effect", and the changes in compliant behavior patterns that are linked to adaptation behavior to stress (habituation vs. sensitization; Herman, 2013 \cite{herman2013neural})become visible. Whereas in the beginning heavy media consumption helps to cope with the stressful situation and translates in momentary increase in trust in media and politics agreement over the time media consumption decreases as well as trust. Hence, it is necessary to investigate the temporal development of biases in news coverage to gain better understanding  of relations between media bias and behaviors related to the COVID-19 pandemic \cite{mehrabi2021survey}.

Moreover, the role of "filter bubbles" \cite{pariser2011filter} shows in first publications covering the  COVID-19 pandemic in terms of polarization and politization of news coverage of the virus. For instance, Boberg et al. (2021) \cite{boberg2020pandemic} used computational content analysis to investigate ``fake news'', which were shown to increase fears, spread opposing views, and kindle a distrusting worldview. Also, in the research on polarization Imhoff \& Lamberty, 2020 \cite{imhoff2020bioweapon} the role of fake news was explored. Politization was shown by Hubner (2021) \cite{hubner2021did} through research of combined computational and manual coding to track virus coverage in the first two months of the pandemic. This study showed politicization related to failures in the communication of scientific facts. This failure in communicating facts was attributed to being a driver of a more political perception of the COVID-19 pandemic.

\section{Case Study} \label{case-study}

This study investigates the effect of a global crisis, i.e., the COVID-19 pandemic, on online news content and its readers' behavior. 
We explore and compare the impact of data-sets collected before and during the pandemic on session-based recommendation models, specifically on BERT4Rec \cite{10.1145/3357384.3357895} - a sequential recommendation model.
In contrast to previous work, we differentiate between the COVID-19 waves. 
First, we conduct an exploratory data analysis and investigate: 

\newcommand{\Qone}{\noindent\textbf{Q1} \emph{How does the COVID-19 pandemic affect the content of a news outlet?\\}}
\Qone
\newcommand{\Qtwo}{\noindent\textbf{Q2} \emph{How does the COVID-19 pandemic affect online newsreaders' commenting behavior?\\}}
\Qtwo
In our second line of research, we train and evaluate BERT4Rec within different stages of the COVID-19 pandemic and study:

\newcommand{\Qthree}{\noindent\textbf{Q3} \emph{To what extent are disparities in data sets (collected during different stages of the COVID-19 pandemic) reflected on the recommendation performance of a sequence prediction model trained with them?\\}}
\Qthree
\newcommand{\Qfour}{\noindent\textbf{Q4} \emph{To what extent do those data set differences affect the topical diversity of the trained recommendation models?\\}}
\Qfour
To answer the stated questions, we conduct our experiments upon a large real-world anonymized data set provided by STANDARD - an Austrian news outlet with the biggest online community in Austria.
The core data set contains articles, user postings, click-streams, and other metadata - collected from 1999 to 2021. For each line of research we follow, we detail our sampling strategy and the features we employ.

\subsection{Analyzing the Temporal Dynamics of Topics} \label{analyzing-the-temporal-dynamics-of-topics} \label{section:content-based-perspective}
In this section, we analyze the temporal dynamics of the topics covered by STANDARD before and during different waves of the COVID-19 pandemic.
Moreover, we also explore the temporal dynamics of the users' posting behavior, i.e., the change under which content people prefer to join the online discourses.
We distinguish between following four different stages of the COVID-19 pandemic: ``pre COVID-19'' (2019-01-01 - 2019-12-30), ``1st wave'' (2019-12-31 - 2020-06-15), ``2nd wave'' (2020-06-16 - 2021-02-10) and ``3rd wave'' (2021-02-11 - 2021-07-03).
The intervals are defined based on Austria's daily COVID-19 cases\footnote{https://commons.wikimedia.org/wiki/File:Coronawellen\_und\_Toteszahlen\_in\_Österreich.svg}.
For each stage, we retrieve all articles (around 142K) and their corresponding keywords.
Note that keywords are defined and provided by STANDARD.
Additionally, for each article, we retrieve its corresponding user postings (around 42M). 

\Qone
Table \ref{tab:top5keywords} shows the top5 keywords assigned to articles over the four COVID-19 stages in Austria.
In the pre-COVID-19 stage, the leading topic is the automobile (vehicles).
However, also other topics reach the top5, like Brexit and the election campaign.
In the first wave, the COVID-19 pandemic is the leading topic, and related keywords appear more than two times more often than other keywords reaching the top5 (automobile and election campaign).
In the second wave, the COVID-19-related keywords overtake the top5 keyword list.
Finally, in the third wave, keywords related to the pandemic are still dominating, but other topics, such as the automobile, begin to reappear.
Moreover, there are, in general, fewer keyword occurrences compared to the other stages. 

\begin{table}[t]
\caption{Top5 Keywords per Article and Their Change over Time Before and During the Three Covid-19 Waves in Austria}
\label{tab:top5keywords}
\begin{tabular}{@{}llll@{}}
\toprule
pre COVID-19 & 1st wave & 2nd wave & 3rd wave \\ \midrule
Auto (2203) & Coronavirus (3104) & Coronavirus (2937) & Coronavirus (1289) \\
Wahlkampf (1484) & Auto (2975) & Corona-Krise (2223) & Krise (880) \\
Brexit (1401) & Corona-Krise (2506) & COVID-19 (1985) & Auto (744) \\
Krise (992) & Wahlkampf (1750) & Krise (1759) & COVID-19 (695) \\
Fahrzeug (992) & Virus (1606) & Virus (1368) & Corona-Krise (664) \\ \bottomrule
\end{tabular}
\end{table}

\begin{figure}[t]
  \centering
  \includegraphics[width=0.9\linewidth]{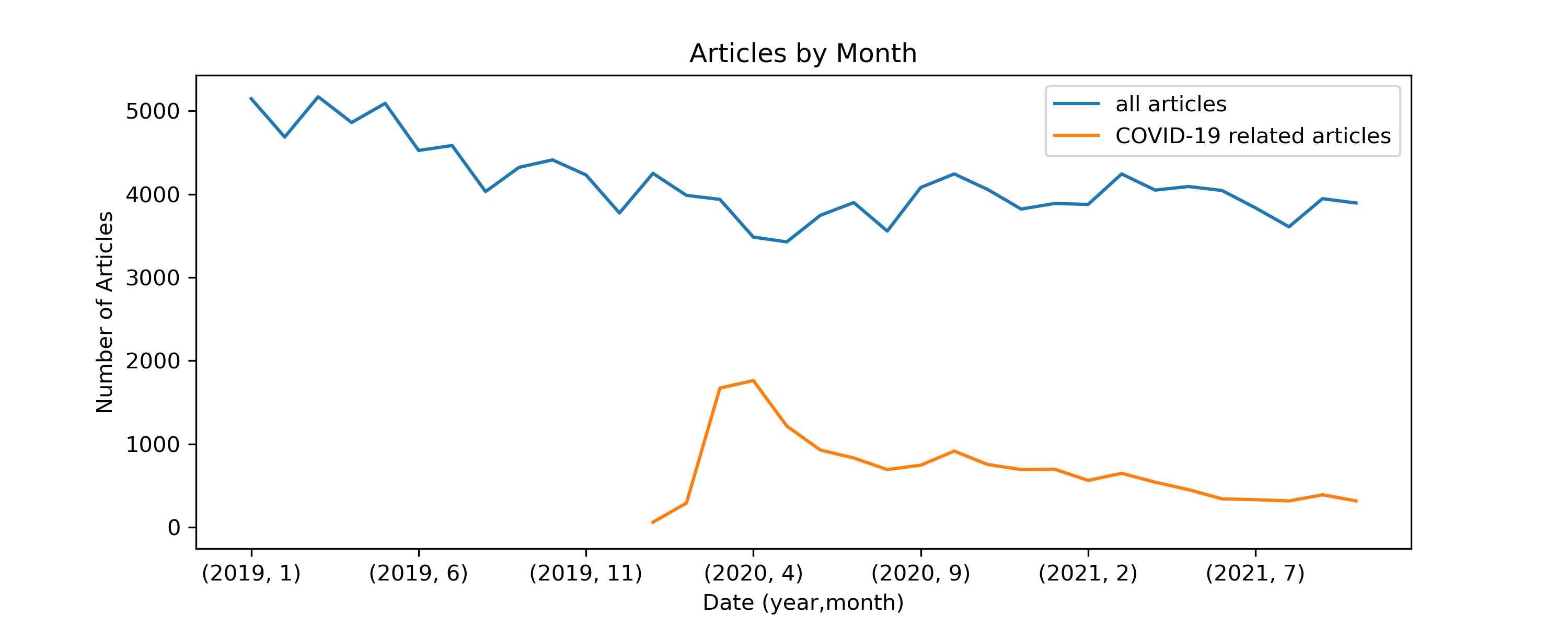}
  \caption{Articles by Month}
  \Description{Articles by Month}
  \label{fig:articles_by_month}
\end{figure}

\begin{figure}[t]
  \centering
  \includegraphics[width=0.9\linewidth]{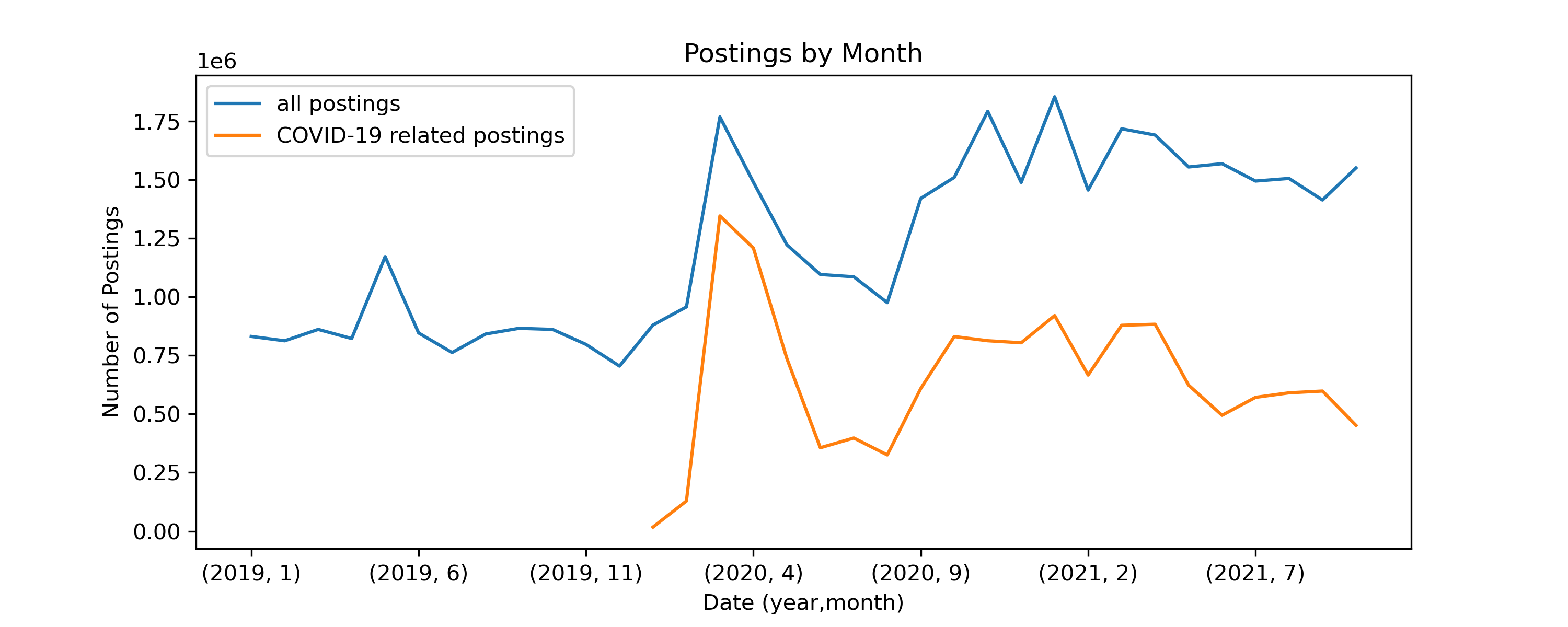}
  \caption{User Postings by Month}
  \Description{User Postings by Month}
  \label{fig:user_postings_by_month}
\end{figure}

Figure \ref{fig:articles_by_month}) provides a more granular picture of the temporal dynamics of the frequency of published COVID-19-related articles.
We plot the number of daily published articles aggregated by month and distinguish between all and only COVID-19-related articles. 
In general, there is a slight decrease in overall published articles in the pre-COVID-19 phase, which more or less stabilizes afterward. 
In contrast, the amount of COVID-19-related articles spikes during the first wave and then steadily decreases.

\Qtwo
Analog to the number of published articles, we also explored the number of user postings below those articles, which we illustrate in Figure~\ref{fig:articles_by_month}).
While the number of all postings is stable in the pre-COVID-19 phase, there is a substantial increase during the first COVID-19 wave, which is overall sustained over the next two waves.
In contrast, postings under COVID-19-related articles spike during the first wave and slightly decay afterward (in line with the number of COVID-19-related articles published - see Figure~\ref{fig:user_postings_by_month}). These temporal effects are in line with the original conceptions and earlier evidence of event related "news bursts" and ``rally around the flag'' effect. Thus, it mirrors the interplay between media coverage and behavior to stressful events.

%The waves are defined as: ``pre COVID-19'' (2019-01-01 - 2019-12-30), ``1st wave'' (2019-12-31 - 2020-06-15), ``2nd wave'' (2020-06-16 - 2021-02-10) and ``3rd wave'' (2021-02-11 - 2021-07-03). The intervals were defined based on the daily COVID-19 cases of wikipedia\footnote{https://commons.wikimedia.org/wiki/File:Coronawellen\_und\_Toteszahlen\_in\_Österreich.svg}. This illustrates the major impact of COVID-19 on news coverage. 

% \usepackage{booktabs}

% \subsubsection{Preprocessing}

% The in table \ref{tab:top5keywords} shown change in topics ...

% \begin{itemize}
% \item 141908 Articles from 2019-01-01 to 2021-09-11
% \item 15259 related to corona
% \item 41667639 user postings from 2019-01-01 to 2021-09-11
% \item 14250548 user postings related to corona
% \end{itemize}

% \subsubsection{Results}

\subsection{Analyzing the Impact of Changed User Behavior on Recommender Systems} \label{section:user-behavior}

%To demonstrate the influence of changed user behaviour on a recommender system, the BERT4Rec \cite{10.1145/3357384.3357895} model is utilized. BERT4Rec is a model to perform sequential recommendations.
%Due to the high amount of user interaction data the to be analyzed time range is limited to one week for each stage.
We investigate the impact of the users' behavioral change on a recommender system by employing BERT4Rec as a representative state-of-the-art model. In a future iteration, it is planned to compare several models from different families.
Due to resource-intensive training of BERT4Rec and limited resources, we subsample one-week intervals out of the previously defined COVID-19 stages.
Thus, the following ranges are defined: ``pre COVID-19'' (2019-11-07 - 2019-11-13), ``1st wave'' (2020-03-22 - 2020-03-28), ``2nd wave'' (2020-11-07 - 2020-11-13) and ``3rd wave'' (2021-03-26 - 2021-04-01).
Each stage is the week before the peak of COVID-19 cases in the respective wave.
To select a comparable time range before the COVID-19 pandemic, one week exactly a year before the second wave is chosen.

%As we are analyzing the change of user behaviour on the COVID-19 pandemic the click stream data are limited to articles on the ``derstandard.at'' website. For the sequence prediction a user is defined as one session whereby a session is a chain of interactions. Sessions with more than 50 article interactions where shortened to the recent 50 interactions. The data set contains cookies which do have a longer lasting lifespan and are assigned to sessions. All interactions assigned to a cookie with less than 5 user-item interactions in total were removed to reduce the noise level.
We utilize the click-streams, i.e., sequential user-article interactions, that emerged during the previously defined intervals.
We only consider readers with more than five clicks during the defined periods to filter inactive users.
Furthermore, we limit the session length to the recent 50 clicks.
We employ the RecBole \cite{recbole1.0} library, which incorporates various recommendation models, such as BERT4Rec, and fosters reproducibility. 
We use the default hyperparameters\footnote{https://www.recbole.io/docs/user\_guide/model/sequential/bert4rec.html} defined by the BERT4Rec \cite{10.1145/3357384.3357895} authors.

%o show to what extent such a recommender system is influenced by disparities in data sets the recommendation library RecBole \cite{recbole1.0} with the model BERT4Rec is used. This library allows the use different recommendation models by their default parameters as presented by the original researches of the different models in this case the BERT4Rec publication \cite{10.1145/3357384.3357895}.

% 1st wave als Bezugspunkt

\textbf{Training and Evaluation:}
For each of the defined phases, we split the extracted samples into training- (80\%), validation- (10\%), and test-set (10\%). 
We optimize for MRR@10 and use early-stopping with a maximum of 300 epochs. %Since we employ the default BERT4Rec settings,
A detailed config/hyperparameter setting is provided in the model documentation\footnote{https://recbole.io/docs/user\_guide/model/sequential/bert4rec.html}. 
%For each stage the data set split: 80\% training, 10\% validation and 10\% test is used. As evaluation metric during training the MRR@10 is selected. Training is performed for 300 epochs whereby an validation based early stopping is used. The ``reproducibility'' option of the RecBole framework is selected to enable reproducibility of the results. Since the default settings are used further details regarding the configuration settings can be looked up at the model documentation\footnote{\urlhttps{https://recbole.io/docs/user\_guide/model/sequential/bert4rec.html}}.

\textbf{Change in User Behavior:} Section \ref{analyzing-the-temporal-dynamics-of-topics} shows, that the COVID-19 pandemic had an major impact on the number of published news media articles and user postings during the observed time spans. Another aspect is the change in user behaviour during the pandemic. Table \ref{tab:standard-behaviour-data-set-staticstics} illustrates that the overall traffic on the website ``derstandard.at'' increased by several hundred percent and declined after the 1st wave slowly by keeping a high level. In addition nearly all clicks during the 1st wave were related to COVID-19 related articles. It can also be observed that the number of COVID-19-related clicks decreased after the first wave, while also remaining at a high level. Even in the third wave, almost half of all clicks in the observed week were COVID-19 related. But not only the click behavior changed during the pandemic also the amount of sessions increased by 60\%. The increased number of sessions, reduced number of average actions by session and reduced number of articles by additionally having and increased avg actions by articles indicate that clustering around a smaller number of items occurred during the first wave took place.\\

\begin{table}[]
\caption{STANDARD Behaviour Data Set Clicks on Articles (one week per stage / baseline: pre COVID-19)}
\label{tab:standard-behaviour-data-set-staticstics}
\begin{tabular}{lllll}
\toprule
                                  & \textbf{pre COVID-19} & \textbf{1st wave} & \textbf{2nd wave} & \textbf{3rd wave} \\ \midrule
\textbf{total clicks on articles} & 13 661 853              & \textbf{54 668 121}          & 36 368 412          & 26 076 694          \\
\textbf{covid related clicks}     & 127                   & \textbf{49 815 670}          & 15 951 442          & 13 178 292          \\ \bottomrule
\end{tabular}
\end{table}

\begin{table}[]
\caption{STANDARD Behaviour Data Set Statistics (one week per stage)}
\label{tab:standard-behaviour-data-set-staticstics}
\begin{tabular}{lllll}
\toprule
 & \textbf{pre COVID-19} & \textbf{1st wave} & \textbf{2nd wave} & \textbf{3rd wave} \\ \midrule
\textbf{sessions} & 1 161 437 & 1 851 518 & \textbf{1 967 872} & 1 605 929 \\
\textbf{avg actions by session} & \textbf{7.484} & 6.526 & 7.129 & 7.339 \\
\textbf{articles} & 50 132 & 41 607 & \textbf{50 663} & 44 246 \\
\textbf{avg actions by article} & 173.390 & \textbf{290.408} & 276.915 & 266.376 \\
\textbf{number of intersections} & 8 692 230 & 12 082 717 & \textbf{14 029 075} & 11 785 825 \\
\textbf{sparsity} & 99.9850\% & 99.9843\% & 99.9859\% & 99.9834\% \\ \bottomrule
\end{tabular}
\end{table}

% Topic Diversity auf Basis der Keywords
%% TOP10 Recommendation List für jede Session mit den zugehörigen Keywords der Artikel
%%% 1. Intersection over Union
%%%% https://stackoverflow.com/a/62088819 (all combinations)
%%%% https://studymachinelearning.com/jaccard-similarity-text-similarity-metric-in-nlp/#:~:text=Jaccard%20Similarity%20is%20also%20known,are%20exist%20over%20total%20words.
%%%% Jaccquard similarity
%%% 2. COVID-Related Keywords in der Recommendation List zählen und durch die Anzahl aler Keywords in der Liste dividieren

\begin{table}[]
\caption{User Behavior (Clickstreams) over Time Before and During the Three Covid-19 Waves in Austria by Performing Sequence Prediction with BERT4Rec}
\label{tab:clickstream}
\begin{tabular}{lllll}
\toprule
 & \textbf{pre COVID-19} & \textbf{1st wave} & \textbf{2nd wave} & \textbf{3rd wave} \\ \midrule
\textbf{mrr@10} & 0.1679 & 0.2414 & 0.2512 & \textbf{0.2575} \\
\textbf{hit@10} & 0.2771 & 0.3742 & 0.379 & \textbf{0.3816} \\ \bottomrule
\end{tabular}
\end{table}

\begin{table}[]
\caption{Topic Similarity Based on Recommended Article Keywords}
\label{tab:similarity-based-on-the-recommended-article-keywords}
\begin{tabular}{lllll}
\toprule
                                  & \textbf{pre COVID-19} & \textbf{1st wave} & \textbf{2nd wave} & \textbf{3rd wave} \\ \midrule
\textbf{Jaccard similarity} & 0.00577              & 0.07986          & 0.01685          & 0.01384          \\ \bottomrule
\end{tabular}
\end{table}

\Qthree Table \ref{tab:clickstream} presents the results after training and evaluating the BERT4Rec models. It can be seen that there is a certain correlation between the change in news articles and posting/click behaviour compared with the mrr@10 \& hit@10 metric. Especially if one compares the pre COVID-19 timespan with the three waves. Interestingly, both metrics remained at high levels after the first wave, although the number of clicks/sessions and COVID-19-related articles/posts declined.

\Qfour
%To make a potential recommendation bias towards certain topics visible, the pairwise similarity of the top10 recommended articles for each session in the test data set is computed. For this purpose, Jaccard similarity or also known as intersection over union (see equation \ref{formula:jaccar-index}, whereby here ``A'' = topics of article one and ``B'' = topics of article two) is used to calculate the similarity of the article pairs and their corresponding topics.
We compute the pairwise Jaccard similarity of the top10 recommended articles for each session in our test splits. 
Therefore, we take the intersection-over-union of the keywords occurring in the document pairs.
This results in a value between 0 and 1, where 1 means maximal similarity. 
The results are aggregated and averaged, resulting in a recommendation similarity metric.
%The results are aggregated and averaged
%, resulting in a recommendation similarity metric for each phase of the pandemic.
Table \ref{tab:similarity-based-on-the-recommended-article-keywords} shows that the recommender system recommended articles with similar topics more frequently during the pandemic than before the pandemic.

% \begin{equation} \label{formula:jaccar-index}
% J(A,B) = \frac{ \left\lvert A \cap B \right\rvert }{ A \cup B }
% \end{equation}

\section{Discussion}
%This section summarizes the issues raised and described in the case study (see section \ref{case-study}) and answers them based on the results obtained by (1) analyzing the temporal dynamics of topics and (2) analyzing the impact of bias on the perception of the pandemic.
First observations derived from the descriptive analysis, mirror assumptions on media coverage and user behavior as described in earlier research on COVID-19 and the media with much smaller data sets. The quantity of news articles associated with "news bursts", the emerging patterns found in the coverage from STANDARD do indeed grow in unprecedented size. The platform contains more than 40 million posts and over 190 million votes from 2019 to the end of 2021. The distribution of the posts over time is shown in figure (\ref{fig:user_postings_by_month}). Each of the articles is tagged with multiple keywords describing the topics contained in the article. Regarding our  question of role of bias in news recommendation in the perception of the pandemic, our exploration showed several patterns as predicted:  corresponding with the rapid increase of media coverage (within the first month of the detection of the first COVID-19 case the numbers of articles increased ten-times), articles on the virus dispersed in most news channels very fast. One signifier of this development, is the key word crisis which was added to the more neutral descriptor of coronavirus already in March 2022. 

Regarding the ``rally around the flag effect'', we could show the increase in user behavior in the assumed the direction of ``rally around the flag'' effect in which showed a decline in activity over time after an initial increase. Moreover, it can be assumed that this change in user behavior also follows pattern as described within the literature on the ``rally around the flag'' effect where the increase in trust in governments in the beginning of a crisis quickly fade in line with earlier research on this effect \cite{johansson2021rally} and so did the traffic wane with the waves. In addition, other studies on the Austrian context on trust in government and vaccines underline this relation between distrust in the government and hesitancy towards measures, especially vaccinations \cite{schernhammer2022correlates}. From a psychological perspective, adaption behaviors to stressful situation can also explain changes in user behavior. These adaptation behaviors to stress (habituation vs. sensitization; Herman, 2013 \cite{herman2013neural}) show in different reactions by individuals confronted with stressful situations. Whereas some individuals become less alert and thus also possible less inclined to engage with a topic, other people show an increase in stress reactions. Both, habituation and sensitization might lead to decrease in compliance with measures, hence albeit being cognitive valuable coping strategies, responsible NRs could help to prevent over stimulation altogether. Hence, it could be counteracted by responsible recommendation that break the ``filter bubble'' for instance by an increase in diversity of recommended articles which allows breaks from disturbing news which is supposed to help to maintain compliance with measures over longer time periods as neither habituation nor sensitization needs to occur.  

Through sequence  prediction it could be shown that over time the average user history indicates toward the fostering of a problematic ``filter bubble'' effect \cite{10.1145/3437963.3441665,abdollahpouri2021user}. Disparities in data sets (collected during different stages of the COVID-19 pandemic) reflected on the recommendation performance of a BERT4Rec model trained with them and affected the topical diversity negatively. These results are a further indicator of the role on media regarding political polarization and a politization of the virus. Moreover, the responsibility of media outlets to provide more diverse recommendations to prevent disengagement from public discourse is shown. For instance, a study on the Austrian situation concerning COVID-19 vaccines and trust in government showed that political polarization fueled a decline in compliance with COVID-19 measures, decrease in trust in authorities and kindled hesitancy towards vaccines \cite{schernhammer2022correlates}. Considering the national context, we believe it is worthwhile to highlight that Austria, together with Germany and Switzerland was among the European countries with the highest rates of
distrust against government measures with demonstrations fueled by conspiracy theories. In addition to insights generated from our data,  special relevance in responsible NRs can also be assumed given the role of social media on political polarization. Social media outlets have been shown in the above mentionned study of the Austrian context by Schernhammer et al. (2022, \cite{schernhammer2022correlates}) to be more frequently used as main source of information amongst individuals who showed more hesitance towards vaccine related news. Given the potential of larger news outlets to provide more diverse recommendations and thus to counter fake news or information which is deliberately biased, their role is crucial in rebuilding trust and engagement with wider parts of the population. Thus, NRs need to be improved in ways that counter developments as shown in our study where a decline of diversity because similar topics are recommended more often than it was the case before the pandemic became visible. Hence, when aiming to prevent the future deepening of societal divides, the role of responsible NRs in traditional media in contrast to social media in the role of the change in perception of the COVID-19 pandemic, merits further investigation.
\section{Conclusions and Future Work}

To summarize, this study aimed to explore patterns of the role of biases in relation to phenomena as described in the social sciences via machine learning interventions. It explored the relationship between news coverage (topics), user behavior (traffic) and examined via development of a model (training) bias in recommendation of temporal aspects of the COVID-19 pandemic and its encompassing of all life domains. Analyses of live data and the results of training models before and between different COVID-19 waves are presented. This study aims to build a ground for further investigation on how machine learning can impact changes the perception of events and that bias may lead to the creation of "filter bubbles" and societal polarization; hence unsolicited outcomes of human-computer interaction. 

The contribution of this ongoing work, lies mainly in  analysis of large data sets capturing the developments from a health crisis to a multi-faceted crisis over different COVID-19 waves. Despite the growing understanding of the role of media bias on user behavior during the COVID-19 pandemic such temporal analyses were missing. 

The extensive public discourse about filter bubbles, echo chambers, fake news, micro-targeting, and related web-based phenomena during recent elections and referendums shows the strong societal interest in this topic. Future studies would benefit from aiming to go beyond accuracy models. A bridge between computer science and psychological concepts could help to disentangle the role of technological possibilities, media-related responses and human vulnerabilities which nowadays often lead to a toxic mix driven by media-coverage and online discussions. Given the volume, velocity, variety, and veracity of comments on news sites, the potential to provide understanding of the dynamics of user networks and discussion could be harnessed by performing large-scale analysis driven by theory and methodology from communication science and psychology. Another venue for future research could lie in the comparison of other news sources with different editorial agendas and in regard to different media formats (e.g., video versus text), both regarding possible bias in the news coverage and user behavior. Therefore, future work could deepen the work on accuracy and different beyond-accuracy objectives and their dynamics in ways to build news recommendation which helps to foster the cognitive and emotional capacities of news consumers instead of further polarization. 

\begin{acks}
This research is supported by the Christian Doppler Research Association (CDG).
\end{acks}
%\newpage

%\section{Appendices}

\bibliographystyle{ACM-Reference-Format}
\bibliography{sample-base}

\appendix

\end{document}